\documentclass[12pt]{article}

\usepackage{graphicx}
\usepackage{amsmath}
\usepackage{amssymb}

\usepackage{lineno}
\usepackage{caption}
\usepackage{subcaption}
\usepackage{authblk}

\title{Social Sensing of Floods in the UK}

\author{Rudy Arthur$^{1*}$,Chris A. Boulton$^{2}$,Humphrey Shotton$^{1}$,Hywel T.P. Williams$^{1}$}

\date{%
	\textit{
    $^1$Computer Science, CEMPS, University of Exeter, Laver Building, North Park Road, Exeter, EX4 4QE, UK\\%
    $^2$Earth System Science, CLES, University of Exeter, Laver Building, North Park Road, Exeter, EX4 4QE, UK\\%
	}    
    $^*$R.Arthur@exeter.ac.uk\\ \vspace{0.05\textheight}%
    \today
}

\begin{document}

\maketitle

\begin{abstract}
``Social sensing'' is a form of crowd-sourcing that involves systematic analysis of digital communications to detect real-world events. Here we consider the use of social sensing for observing natural hazards. In particular, we present a case study that uses data from a popular social media platform (Twitter) to detect and locate flood events in the UK. In order to improve data quality we apply a number of filters (timezone, simple text filters and a naive Bayes `relevance' filter) to the data. We then use place names in the user profile and message text to infer the location of the tweets. These two steps remove most of the irrelevant tweets and yield orders of magnitude more located tweets than we have by relying on geo-tagged data. We demonstrate that high resolution social sensing of floods is feasible and we can produce high-quality historical and real-time maps of floods using Twitter.
\end{abstract}

\section*{Introduction}
Natural hazards such as floods, wildfires, storms and other extreme
weather events cause substantial disruption to human activity and
are predicted to increase in frequency and severity as the climate
changes \cite{ipcc:2012,muller:2015}. Rapid high-resolution observations of unfolding hazard events can inform decision-making and management responses. However observations from meteorological and climatological instrumentation typically suffer from data sparsity, time delays and high costs \cite{muller:2015}. Furthermore,
the human impacts associated with natural hazards are hard to measure using sensor platforms designed to observe the hazards themselves. Rain gauges and stream gauges can only measure the amount of precipitation, or the height of floodwater, not the impact on people's lives.  Crowd-sourcing has been proposed as a possible solution to the need
for better observations of environmental phenomena and associated
impacts, utilising a variety of methods including citizen observations,
web technologies, distributed sensor networks, and smart devices \cite{muller:2015}.

A useful distinction may
be drawn between ``solicited'' 
crowd-sourcing (where users are
somehow recruited to participate in structured observation programmes)
and ``unsolicited'' crowdsourcing
(where observations are derived
as a by-product of social communication for other purposes). 
Most citizen science and crowdsourcing studies to date have generated solicited
observations \cite{muller:2015}. Examples relevant to natural hazards
include the UK Met Office ``Weather Observation Website'' \cite{wow},
which allows amateur meteorologists to upload local weather reports
through a web application, the European Severe Weather Database \cite{ESWD} which uses eye-witness reports to map severe weather across Europe, and the UKSnowMap \cite{uksnow} initiative that
uses a particular hashtag to allow users to contribute structured
snowfall observations via social media. Solicited crowd-sourcing tools
typically use a designed interface to impose structure on the data
that is collected, which may improve data quality by standardising
the citizen observations obtained. However, such tools rely on dedicated
volunteers choosing to upload data and the volume of observations
that is collected may therefore be limited. 

Social sensing is here defined as observation of real-world events
using unsolicited content from digital communications (e.g. mobile phone call records, social media, web searches, and other online data).
The core challenge of social sensing is to extract high-quality observational
data from large numbers of unstructured, patchy, and possibly inaccurate
user utterances, or from associated metadata. 
Call data records (CDR) from mobile phone usage have been analysed to reveal patterns
of mobility in response to natural disasters \cite{lu2012} and to predict the 
spread of infectious diseases \cite{wesolowski2012,bengtsson2015}. However, CDR data is hard to access 
due to potential commercial sensitivity and privacy concerns. Web search
statistics have also been used to track infectious diseases,  
but are affected by changes in underlying (proprietary) search algorithms 
which may change outputs and confound statistical analyses \cite{lazer2014}.
In this study we consider social sensing using publicly available social 
media data.

Several studies have used social media for 
sensing natural hazards. Data from Twitter has been used for detection/location
of earthquakes \cite{sakaki:2010}, typhoons \cite{sakaki:2010},
wildfires \cite{boulton:2016} and heat waves \cite{Kirilenko:2015}.
Data from other platforms has also been used, including air quality
prediction in China using data from Sina Weibo \cite{Jiang:2015}
and flood prediction using Flickr \cite{Tkachenko:2017}.

The early work of Sakaki et. al. \cite{sakaki:2010} proposed a classifier based on keywords in a tweet, the number of words and their context to produce a probabilistic model for earthquakes that located their centre and predicted the trajectory of the shock. They managed to predict 96\% of significant earthquakes in Japan and moreover were able to produce warnings faster than the Japan Meteorological Agency. This methodology also worked well for typhoons and demonstrated that information on Twitter could be of immense benefit for studying natural hazards. Boulton et. al. \cite{boulton:2016} showed that Twitter could be used to detect wildfires in the US, with tweets about fires showing strong spatio-temporal correlation with satellite data. Similarly Kirilenko et. al. \cite{Kirilenko:2015} showed that Twitter users identify heat waves. The work of Sakaki et. al. was influential, however most studies using Twitter to detect events have focused on epidemics e.g. \cite{Signorini:2011} or political events \cite{Hermida:2014} rather than natural hazards {\it per se}. The field of natural hazard monitoring using Twitter remains fairly under-studied, and certainly many meteorological and disaster management agencies who could benefit from this data are not currently using it.

The advantage of social media data is volume. For example, the popular
social media platform Twitter has over 320 million active users each
month producing over 500 million tweets each day \cite{leading:2015}.
The accessibility and real-time information dissemination
capabilities of Twitter make it a good candidate for social sensing
in the geographical areas where it has high usage. However, Twitter
has an important limitation in that only a small fraction (typically
\textless1\%) of tweets have accurate location information in the
form of latitude/longitude coordinates derived from a GPS-capable
mobile device. 
Instagram has a higher proportion of GPS-tagged content (e.g. $\sim$12\%
of posts were geotagged in a social sensing study of wildfires in
the USA \cite{boulton:2016}). Facebook has very high usage but privacy
constraints severely restrict its use for social sensing. 

A general problem for crowd-sourcing of environmental observations
is data quality. The potential for high-volume and high-resolution
observations is balanced against the potential for inconsistent, inaccurate
and subjective reporting. Social media content
suffers from a lack of structure and standardisation in the observations
that are collected, and ``noise'' in the form of irrelevant content
returned by search filters. 
Social media attention to ``routine'' daily weather may
be low, with more attention given to ``newsworthy'' extreme
events. The networked nature of social media
may exacerbate these problems, since users are exposed to each other's
content and are likely to have a relatively high density of online
connections to other users from the same geographical region 
\cite{jurgens:2013}, the independence of observations arising from a particular
region is unknown. Interpersonal influence and social spreading
of content (often mediated by mechanisms designed for this exact purpose,
for example, retweets, re-posts, likes) may amplify or suppress some
kinds of user report and thereby distort observations. 

Here we describe a case study of social sensing of flooding in the
UK using unsolicited data from Twitter.  Since the geographical location of events is important for this application, we use location inference to estimate the geographical origin of tweets that do not contain GPS coordinates, adapting a method \cite{schulz:2013} that utilises a variety of indicators taken from the tweet object
and user profile to infer the location of the user at the time the
tweet was written. 

This paper has several aims. The first is to discover if social sensing can be used to detect flood events as has been demonstrated for other hazards, such as earthquakes and wildfires. Once this has been shown we want to use Twitter to produce data for flood forecasters in order to help validate and improve their forecasts, particularly of minor floods. At present the methods to validate flood forecasts rely on manual searching of national and local news in areas where there has been a flood forecast. It would be useful to forecasters firstly to have an automated method which removed the burden of performing this manual search and secondly to have a method which can detect very localised floods, which may not make the news. Finally, we note that the tweets we collect often contain useful information about transport disruption, the progress of the flood waters and emergency service response. Thus by spatially locating all this information we can provide historical and real-time maps which could be of use to emergency responders, planners, insurance agencies and others interested in measuring the human impact of floods.

In Section \ref{sec:methods} we first describe our method of data collection as well as our validation dataset. Section \ref{sec:coll} describes how the data was collected, Section \ref{sec:filters} describes how we filter the data stream for relevant tweets in the right geographical area. Section \ref{sec:inference} describes our approach to location inference and \ref{sec:event} shows how we detect flood events given relevant, located tweets. Section \ref{sec:res} covers the outputs of social sensing, validation and parameter tuning. We also show an example of a day with an extreme flood event as an example of how social sensing could be used in practice. Section \ref{sec:discuss} offers some simple conclusions and discusses the potential/limitations of social sensing.

\section{Methods}\label{sec:methods}

The overall method has four stages: data collection, content filtering,
location inference and event detection. These are covered in turn
below. Choices are made at each stage which affect the final outcome
of the analysis. We use the performance of the overall system to tune
parameters for the location inference and event detection methods. 

\subsection{Data Collection}

\label{sec:coll}

\subsubsection{Twitter dataset}

\begin{figure}[!ht]
\centering 
\includegraphics[width=0.44\textwidth]{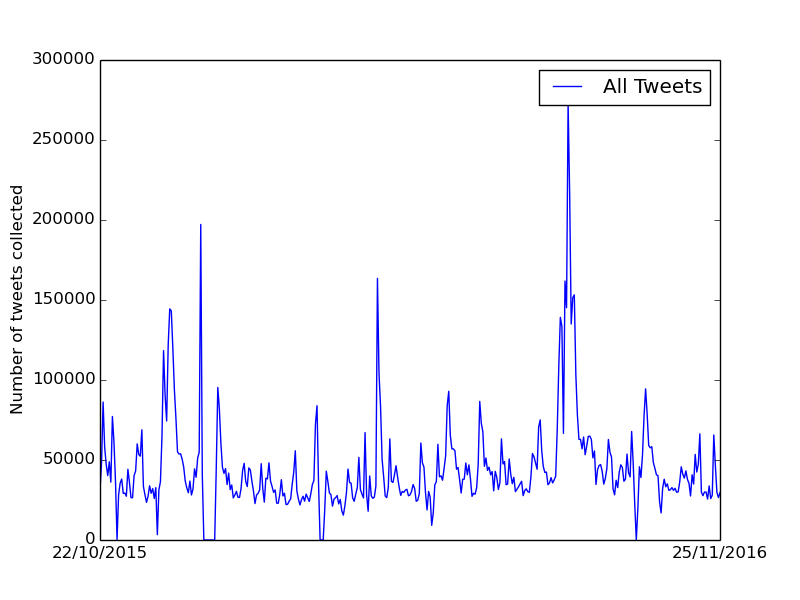}
\caption{ Number of tweets collected per day during the whole collection period 22/10/2015 and 25/11/2016. }
\label{fig:total_count} 
\end{figure}

Tweets were collected from the Twitter Streaming API using the search
terms ``\emph{flood, flooding, flooded}''. All data was collected according to Twitter’s terms of service and privacy conditions. Collection scripts were
implemented in Python using the Twython \cite{twython} 
package. Twitter returns a JSON object for each tweet, this is a common data exchange format consisting of a collection of key-value pairs. The JSON object contains the tweet
content and various meta-data (time stamp, geotag, user profile, etc.). A timeseries of daily tweet counts for this period is shown in Fig ~\ref{fig:total_count}.

\subsubsection{Validation dataset}

We validate our social sensing method against a database of flood observations
collected by the Flood Forecasting Centre. These data record the date,
location (resolved to the level of the UK administrative area in which the flood occurred)
and severity of floods (severe/major/minor) over a period from 01/01/2015 to 23/11/2016.
These data are obtained by manually searching local and national news
for reports of floods and using alternative social media monitoring
tools. This data represents a necessarily incomplete picture of floods
happening in the UK over this period; data are only compiled when
floods have been forecast. There are likely to be few or no ``false
positives'' where floods were recorded but never occurred, but an unknown
number of ``false negatives'' where floods occurred but
were not recorded. We will take these considerations into account
when it comes to tuning the parameters of our algorithm.

\subsection{Filters}

\label{sec:filters}

A top-level filter is instantiated by the search string passed to
the Twitter API (see above). While it is possible to constrain the
search at the level of the API, here we prefer to use a general API
call and filter the returned content post-collection. We use several
additional filters to remove undesirable content, applied in the order
below.

\noindent \textbf{Timezone filter} \newline
The user timezone has been used elsewhere as an indicator for geolocating
tweets \cite{schulz:2013}. For this study the target area
is decided beforehand and tweets by users with timezones set outside
this area are discarded. This has the potential to erroneously remove
some content from users who have unset, or incorrectly set, timezones.
However, checking for wrongly set timezones requires a lot of processing
and would make real-time analysis difficult. Here we require that
users have their timezone set to one of \textit{London, Edinburgh,
UTC.} 

\noindent \textbf{Bot filter} \newline
Some Twitter accounts are automated ``bots'' that tweet a high volume
of flood-related messages. These can often be identified by their
anomalously high level of activity, operationalised here as accounts
whose activity makes up more than $\sim1$\% of the total tweet volume.
We found three such users \textit{@RiverLevelsUK, @UKFloodTweets,
@FloodAlerts}. Since we are interested in \textit{social} sensing we
discard tweets from these users.

\noindent \textbf{Retweet Filter} \newline
Many tweets are users quoting or relaying each other's messages. We
are interested in original and direct flood observations by individuals (e.g. \textit{``Can't
believe my garden is flooded''}) or organisations (e.g. \textit{``We are closed
tonight due to flooding''}), rather than acknowledgements or forwarding
of these observations by others. Thus we filter out retweets.

\noindent \textbf{Relevance filter} \newline
Twitter users often use the search keywords used to collect data (\emph{flood, flooding, flooded})
in contexts that are unrelated to our topic (e.g. \textit{``The market was flooded with copies.''}). To remove irrelevant content, we first remove tweets containing some manually
curated filter terms, 
\begin{verbatim}
"flood with" "flood of" "flood in" "flood it" "flooded with" "flooded by"
"flooding back" "immigrant" "migrant" "migration" "market" "tears" "flood-hit" 
\end{verbatim}
We then apply a machine learning relevance classifier to the remainder. 
To develop the classifier, we first observed that content typically showed 
four identifiable classes of tweets about floods (with some tweets remaining ambiguous): 
\begin{itemize}
\item \textbf{Irrelevant} e.g. \textit{I was in floods of tears} 
\item \textbf{Historical} e.g. \textit{Charity raffle for Cumbrian flood
victims} 
\item \textbf{Warnings} e.g. \textit{River Ouse water level: 2.14m. Chance
of flooding.} 
\item \textbf{Immediate} e.g. \textit{It is flooded outside.} 
\end{itemize}

For forecast verification and flood tracking we want to focus on the
\textbf{Immediate} class: these are the useful social data for detecting
floods currently happening in the UK. We manually curated a set of
$3879$ tweets, classifying each tweet as \textbf{Immediate} or 
\textbf{Other} (including the Irrelevant, Historical and Warnings categories identified above) and used them to train a Naive Bayes \cite{russell:2003} classifier. These tweets were  randomly chosen from tweets which passed the other filters and manually classified. 
To build the classifier we first vectorized the data by counting single-word and 2-gram (two-word) occurrences in the training corpus. We used a multinomial naive Bayes
classifier \cite{scikit} with smoothing parameter $\alpha=0.5$, the exact value of $\alpha$ is not crucial, the results below are very similar for a range of possible smoothing parameter values. To validate, we first split the training data and
find 80\% of tweets classified correctly when the model is built from 75\% of the data. 
Then, using all the data, we perform 6-fold cross validation. We find the sum of the confusion matrices from each of the 6 cross validations is 
$$
\begin{pmatrix}
TN & FP\\
FN & TP
\end{pmatrix}=\begin{pmatrix}
1759 & 359\\
337 & 1360
\end{pmatrix}
$$
These results imply that the classifier is not biased towards false negatives
or false positives.

\subsection{Location inference}\label{sec:inference}

As stated above, typically $<1\%$ tweets have GPS coordinates attached as metadata. 
Location inference is therefore an active topic in analysis of social media.
Our method is based on the multi-indicator approach developed by \cite{schulz:2013},
which uses a variety of indicators to infer tweet origin, including
user timezone, place names mentioned in tweet text, user location field,
GPS coordinates, etc. However, unlike other studies using location inference
\cite{ajao:2015,compton:2014,rahimi:2015,schulz:2013}, here we are
not trying to locate individual users and tweets, but to locate flood
events. Our variant of the location inference problem allows for some
simplification compared to the general problem of locating a given
tweet without any contextual information. 

In order to provide for potential usage for real-time updates, the
location inference process must be sufficiently fast. As we will describe below,
calls to some external databases are needed to identify place names (toponyms) in
tweets. The public interfaces of these services are rate-limited and, since they require user accounts, we cannot perform multiple queries in parallel. DBPedia Spotlight implements an algorithm to detect toponyms, which can take time (seconds) to yield results. These facts can be an issue during active periods, e.g. a large flood in a populous area. This means it is important to only try to locate relevant
tweets, rather than every tweet that is collected, so the filtering described above is of key importance. Since
we suppose that the algorithm developed here will be applied to detect
floods in a specific area (e.g. our study focuses on floods in the
UK) we utilise the user timezone associated with each tweet as a simple
filter prior to location inference, rather than as part of a generalised
location inference procedure.

If a tweet passes through all the filters it is likely (though not
certain) to be relevant, i.e. to concern a flood currently happening
somewhere in the UK. The two most important indicators for locating
the tweet more precisely (assuming the absence of GPS coordinates in metadata) 
are place names mentioned in the tweet text and
place names mentioned in the user location field. Since the flood-focused
tweets analysed here are relatively likely to mention specific places
(i.e. where the flooding is occurring) in comparison to other studies
\cite{schulz:2013,rahimi:2015}, and since we are seeking to locate floods 
rather than user home locations, we can assume that message text is
more likely to contain useful location information than the user location
metadata. For example the tweet text ``\textit{Train from Exeter to Totnes cancelled due to flooding on the
track.''} is very useful for location inference and is more relevant than
the location field for the train company which authored the tweet, which contains ``South-West UK''. This fact will be reflected in the relative weighting
of text and location fields in the inference process.
Since we are ultimately trying to locate floods (not people or tweets),
when tuning parameters and other design options, we focus on
the overall accuracy of flood event detection, rather than the accuracy
of location inference on a single tweet, as the metric for improvement.

\subsubsection{Method}

A small subset of tweets have a \textit{geotag} - precise GPS co-ordinates (from a
GPS unit in a mobile device) of the tweet's origin included in the metadata of the tweet object. For the small subset of geotagged tweets we can simply use the GPS co-ordinates 
and no inference is necessary.  For the majority of
tweets, which are not geo-tagged, we use the method below, based on location information from the user location field and the tweet text. To look up polygons outlining named areas we make use of GADM \cite{GADM}, a spatial database of the location of the world's administrative areas or administrative boundaries.

\textbf{Location Field}: A small number of users have GPS co-ordinates in their
location field (these are independent of any GPS information in tweet metadata), 
usually automatically added by a mobile device. For those users
we use this GPS co-ordinate. The vast majority of users have text in their location field
which we look up in the online gazetteer
Geonames \cite{geonames}, which, if successful,
returns possible locations and a similarity score, indicating confidence in the result. 
If no result is found by looking up the whole location field we split the text
by commas, slashes or hyphens and look up each entity separately. 
Geonames provides the country each location is in. Since we are concerned with floods
in the UK we keep only results from the UK, disambiguating e.g. Boston,
Linconshire from Boston, Massachusetts but not Cambridge, Gloucestershire
from Cambridge, Cambridgeshire. 
If Geonames reports the place as a `region', `area', `city' or `state' 
we look up the location in GADM to obtain a polygon describing the location, otherwise
we use the latitude and longitude provided by Geonames. We store the
polygons or coordinates, together with the similarity score, which we use
as an approximate quality score for each possible location indicated by 
location field. 

\textbf{Message Text}: We use DBPedia Spotlight \cite{isem2013daiber} 
to identify all entities in the whole tweet message text 
and link them to DBPedia entries.
If DBPedia Spotlight tags an entity as a place, we first attempt to
look up the place in GADM. If the place is not found in that database
we follow the link to DBPedia and use the latitude and longitude obtained
from there. DBPedia Spotlight returns a similarity score for each
entity it tags, which we use as a quality score. 

\textbf{Overall Inference:} 
The results of this process are several polygons or coordinate pairs with different quality scores.
For example, for a tweet with a user location field containing \textit{Cumbria
/ London} and the message \textit{Terrible flooding in Carlisle today,}
we would obtain three objects (two polygons for Cumbria and London, and
point coordinates for Carlisle), each with a different quality score. To obtain
the most likely origin of the tweet, taking into account the differing quality 
of each estimated location, we weight each polygon/point by its quality score 
and calculate the sum of the weights of intersecting areas. This example would yield Carlisle, which is
in Cumbria, as the most likely location (i.e. the location with the highest summed weight). 
In this way we obtain a list of the most likely points or polygons associated with each tweet.
It is also possible to skip this step and keep all the polygons associated
with every tweet, but this was found to give slightly worse agreement
with the verification data, so we use the overlap method and keep
only the most likely location. As we mentioned before, the importance
of place names mentioned in the text versus place names in the location
field may be different. We multiply the weights of the message polygons
by a factor $r$, where $r$ will be tuned to optimize agreement with
the verification data.

\subsection{Flood event detection}\label{sec:event}

After filtering and location inference, we are left with a set of relevant 
geo-located tweets which we take as local observations of floods. The next task
is to combine these tweets into a temporally resolved map of flood events for 
comparison with our validation dataset from FFC. 

We add all tweets collected over some period, usually 24
hours, to a map to find the places with the highest volume of flood
related tweets. To this end we start with a spatial bounding box containing
England and Wales (Scotland and Northern Ireland are not included
in our verification dataset) and divide it into an $N\times M$ grid.
Every grid square, $g$, initially has height $g_{h} = 0$. For
every polygon $p$ and grid square $g$ we add a value proportional
to their overlap:
\[
g_{h} \rightarrow g_{h} + \frac{\text{Area of }g\cap p}{\text{Area of }p}
\]
When $p$ is a point location, or entirely inside a grid square,
we add 1 to the grid square containing $p$. This means every tweet
counts equally, but if a tweet is only roughly located e.g. $p$ is a polygon
covering all of Wales, we will add a very small amount to each grid square 
intersecting Wales. This process privileges precise locations, meaning that tweets for which GPS coordinates are available will contribute strongly to flood detection.

Population density must be accounted for since, for example, the total
volume of tweets from London is so large that, based on absolute numbers,
it almost always appears flooded. Ideally we would use the total number of tweets from an area to normalise, however we do not have access to this data. We use the 2011 UK census \cite{census} to obtain very fine grained (though slightly out of date) population density
information which allows us to calculate the population inside each
grid square, $N_{g}$. We then scale the heights 
\[
g_{h}\rightarrow\frac{g_{h}}{N_{g}^{\alpha}}
\]
where $\alpha$ is an adjustable parameter. $\alpha=0$
removes the scaling for population density, while increasing $\alpha$ more severely
down-weights grid squares with large populations and hence
gives more importance to tweets about floods from sparsely populated
areas. We find appropriate values for $\alpha$ by tuning its value
based on overall flood detection performance (see below).


\section{Results}

\label{sec:res}

\subsection{Data Preparation}

\begin{figure}[!ht]
\centering 
\begin{subfigure}{0.44\textwidth} \includegraphics[width=1\textwidth]{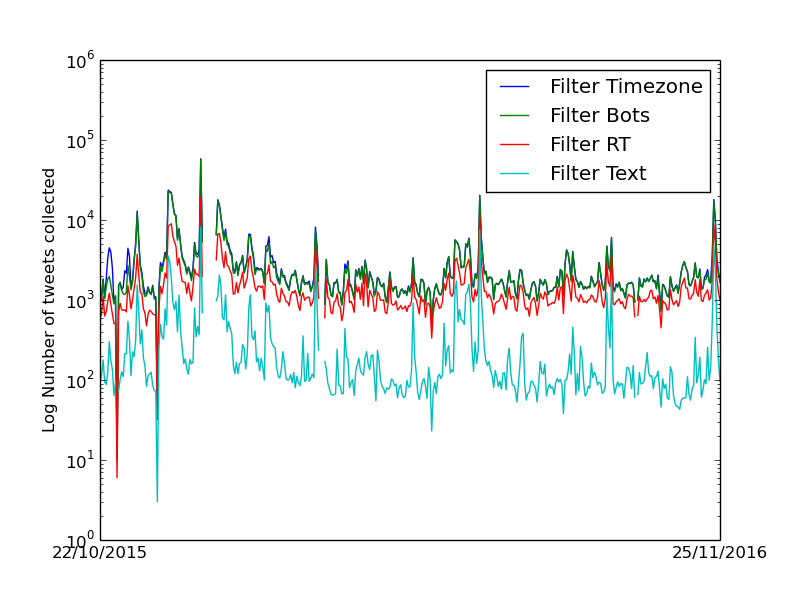}
 \end{subfigure} 
\begin{subfigure}{0.44\textwidth}
\includegraphics[width=\textwidth]{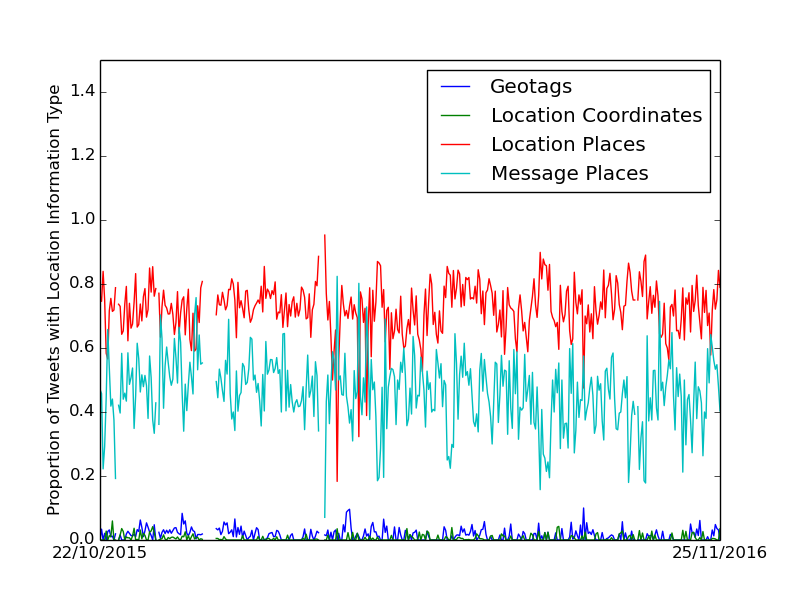}
\end{subfigure} 
\caption{Daily tweet counts with respect to filtering and geographical indicators. Left: Number of tweets collected per day during the whole collection period 22/12/2015 and 04/01/2016 at each filter level. Right: Number of relevant tweets collected with location info in each field: GPS-tagged tweets, location field GPS coordinates, location field toponyms, message text toponyms. }
\label{fig:counts}
\end{figure}

17,828,704 tweets were collected between 22/10/2015 and 25/11/2016 with a gap in collection
between 28/12/2015 and 04/01/2016 inclusive. Table \ref{tab:count} shows the numbers of tweets retained after each filtering step.
Fig \ref{fig:counts} show daily tweet counts broken down 
by filtering and location inference outputs.
In order to get a rough idea of the benefit of each filtering step we measure the Pearson
correlation between daily tweet counts and the FFC validation data. To make the FFC data 
more directly comparable, we create a count for each day by summing the populations of the flooded counties multiplied by the severity of the flood in the county (using a multiplication factor of 1 for a minor flood, 2 for a significant flood and 3 for a severe flood). This construction assumes that the number of tweets is directly proportional to population and that more severe floods generate more tweets. The result of this comparison is shown in the last row of Table \ref{tab:count}. Each of the filters (apart from bot removal which is necessary to remove redundant forecast information) increases the correlation. This general pattern is robust to different transformations of the FFC data, though the amount of correlation changes.
See Table \ref{tab:loccount} for details
of location information associated with the filtered tweets.
Of the 122,281 tweets retained after filtering, 79,163 had some form of 
location information associated with them, with just 1,574 having a geotag.

\begin{table}[!ht]
\centering
\caption{
{\bf Total number of tweets remaining after each filter is applied and correlation of the number of tweets per day with FFC data.}}
\begin{tabular}{ |c|c|c|c|c|c| }
\hline
& All & Timezone & Bot & Retweet & Relevance \\ \hline
Tweets Remaining & 17828704 & 1105360 & 1051295 & 582141 & 122281 \\ \hline
Correlation & 0.206 & 0.551 & 0.550 & 0.591 & 0.673 \\ \hline
\end{tabular}
\label{tab:count}
\end{table}

\begin{table}[!ht]
\centering
\caption{
{\bf Total number of tweets with each kind of location information. }}
\begin{tabular}{ |c|c|c|c|c|c|c| }
\hline
All Tweets & Any Location Info & Geotag & Loc: GPS & Loc: Toponym & Text: Toponym \\ \hline
122281 & 79163 & 1574 & 349 & 59179 & 41315 \\ \hline
\end{tabular}
\begin{flushleft} Tweets can have location information in multiple fields: the Geotag, the location field (Loc), or in the tweet message itself (Text).
\end{flushleft}
\label{tab:loccount}
\end{table}

\subsection{Real Time Flood Maps}

\begin{figure}[!ht]
\centering \begin{subfigure}[t]{0.44\textwidth} \includegraphics[width=1\textwidth]{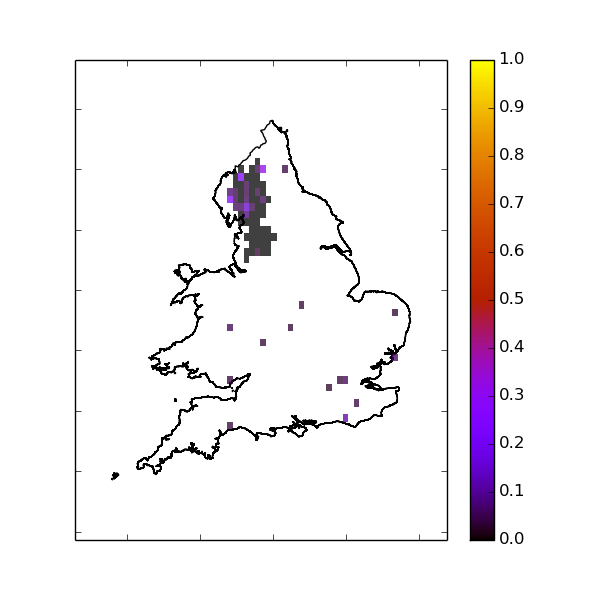}
\end{subfigure} \begin{subfigure}[t]{0.44\textwidth}
\includegraphics[width=1\textwidth]{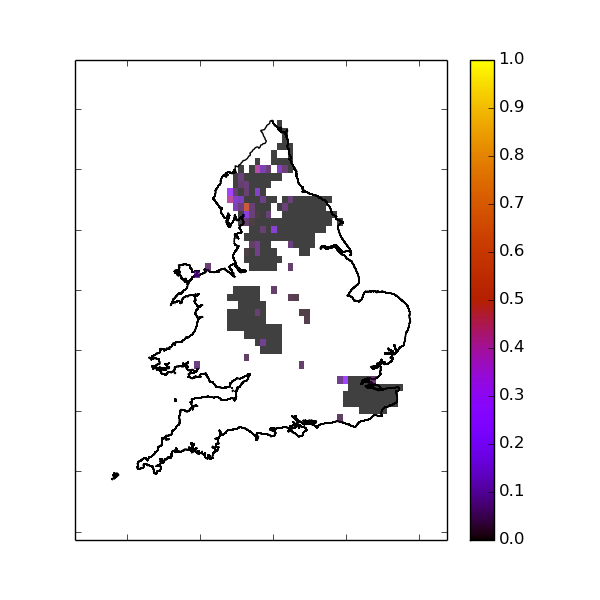}
\end{subfigure} \\
 \begin{subfigure}[t]{0.44\textwidth} \includegraphics[width=1\textwidth]{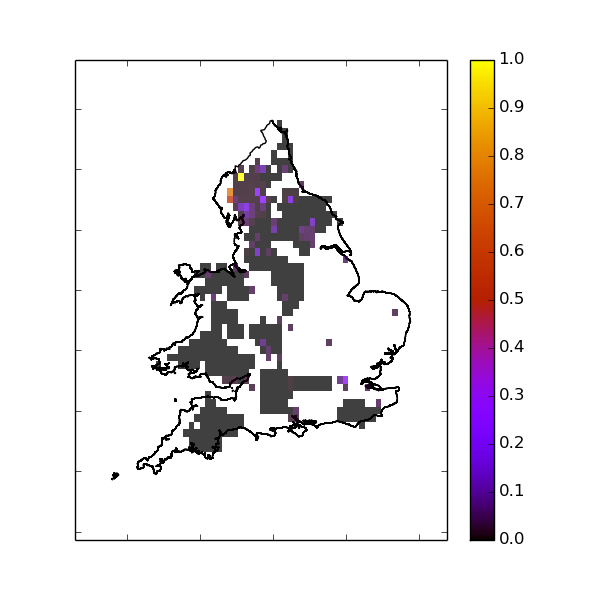}
\end{subfigure} \begin{subfigure}[t]{0.44\textwidth}
\includegraphics[width=1\textwidth]{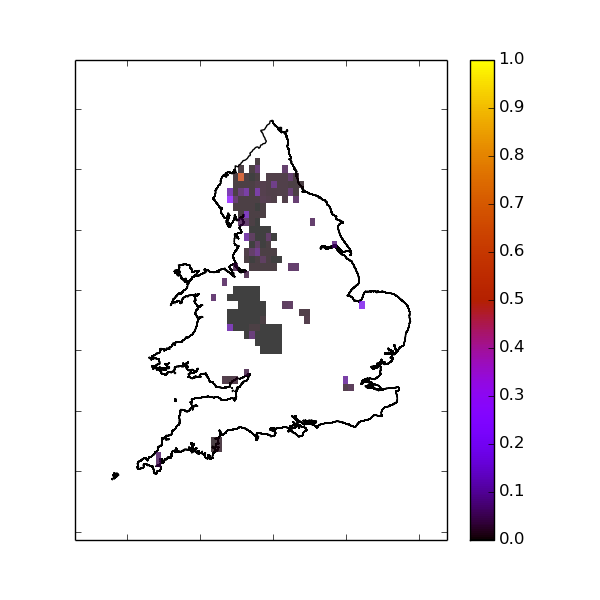}
\end{subfigure} \caption{ Floodiness grid, $64\times64$, over England and Wales on 5/12/2015
using $(r,\alpha,T)=(1.0, 0.15,0.1)$. Using tweets collected in 1 hour windows. White indicates no tweets. Colour bar units are floodiness relative to daily max. Top left: 10am-11am.
Top Right: 1pm-2pm. Bottom Left: 4pm-5pm. Bottom Right: 9pm-10pm. }
\label{fig:days} 
\end{figure}

During large floods there is likely to be a high volume of Twitter
activity referencing the event. Thus we can make `live' maps of Twitter
activity which may be of use to individual travellers, train and bus
companies (though these companies provide a lot of the most useful
tweets) or flood management agencies. Fig \ref{fig:days} shows
an example of four one hour periods on a particularly floody day:
05/12/2015 which saw severe flooding in the north of England.

\begin{figure}[!ht]
\centering \includegraphics[width=0.65\textwidth]{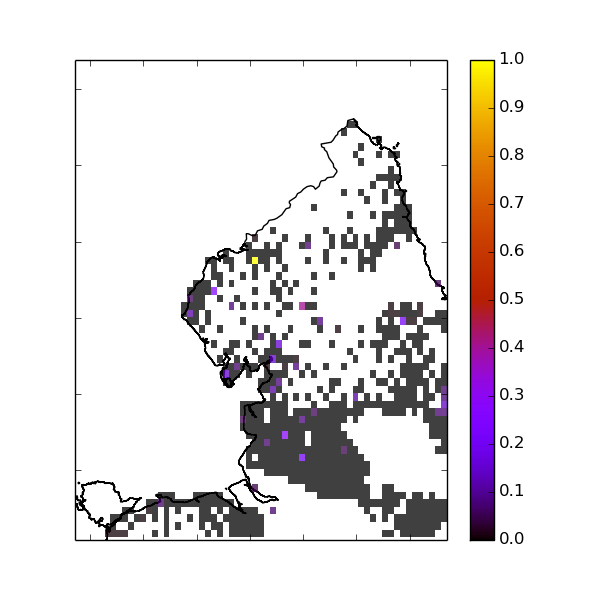}
\caption{ Floodiness grid, $64\times64$, over the North East on 5/12/2015 between 4pm and 5pm using $(r,\alpha,T)=(1.0, 0.15,0.1)$. White indicates no tweets or zero population. Colour bar indicates floodiness relative to daily max.}
\label{fig:zoom} 
\end{figure}

In Fig \ref{fig:days} there are indications of floods early in
the morning. Later in the afternoon we can see flood hotspots
in three places, a town in the area (Kendal) and on the road between
the two major cities in the area Carlisle and Newcastle. In the early evening there is a lot of activity in Carlisle, which continues into the night. If there are sufficiently many tweets the grid can be refined to provide more detailed local information, as in Fig \ref{fig:zoom}.

\subsection{Validation and Tuning}

To validate our method we use the manually curated flood observation
dataset provided by FFC. Since our algorithm detects floods by the volume of tweets observed
in arbitrary grid squares, whereas the FFC system uses UK administrative
areas (typically counties) we need to map our grid data onto the same
geographical regions used by FFC (hereafter counties). There are several
ways to do this, from which we select a method that mimics how
floods are recorded by the FFC. For each observation
period (24 hours from midnight to midnight) we calculate the grid
for some choice of parameters $r$ (relative weighting of message
text and user location field in location inference) and $\alpha$
(population density scaling). We divide by the maximum height
so that the highest square has $g_{h}=1$. We then set a threshold
$T$ and use the criteria: if a grid square overlapping a country
has $g_{h}$ greater than $T$ we declare that county flooded and
check for agreement with the FFC data.

If we call $g_{h}$ `floodiness' then by doing the normalization we
are measuring relative rather than absolute floodiness. This means,
even on a day with no floods anywhere in the UK, some particular point
will be flagged as flooded. Both absolute and relative floodiness are useful metrics and both should be reported. One of the use cases of this method is for forecast verification and flood tracking. In this case the relative measure, which highlights the most flooded areas, will be useful, even if the day is not particularly `floody'. In other situations, e.g. fine days with no floods forecast, the absolute measure is more useful. Indeed, on days with no floods it is unnecessary to try to socially sense them! When we are trying to socially sense floods we will have a good reason to expect a flood somewhere - e.g. one was forecast, heavy rainfall. 

For tuning we chose one day at random from each month where we had both verification data from the FFC and had collected tweets, since days where FFC do not observe floods do not provide the necessary data for validation. We need to span a long time period in order to capture possibly different patterns of twitter use e.g. floods in summer compared to floods in winter may cause different patterns of tweeting.

In the following a \textit{positive} is a county where (for some values
of $r$, $\alpha$) the floodiness was larger than $T$ in some grid
square intersecting that county and a \textit{negative} is a county
intersecting no grid square with floodiness larger than $T$. True
positives (TPs) then are counties socially sensed to be flooded and also recorded
in the FFC data, false positives (FPs) are socially sensed to be flooded
but not in the FFC data. False negatives (FNs) are counties marked by
the FFC but not social sensing.

To tune we choose 14 days where we have both FFC observations and
tweets and count the number of TPs, FPs and FNs for different triples
of $r$, $\alpha$ and $T$ on each day. We then calculate 
\begin{equation*}
\text{Precision }=\frac{TP}{TP+FP} \qquad \text{Recall }=\frac{TP}{TP+FN}
\end{equation*}
for each parameter set for each day and compute average precision
and recall over the 14 days. Using average precision and recall in
this way means that each day contributes equally to the final scores,
whereas compiling all TP/FP/FN scores into a single confusion matrix
for calculation of a single precision and recall value would risk
over-fitting to a single day. 

\begin{figure}[!ht]
\centering 
\begin{subfigure}[t]{0.4\textwidth} 
\includegraphics[width=1\textwidth]{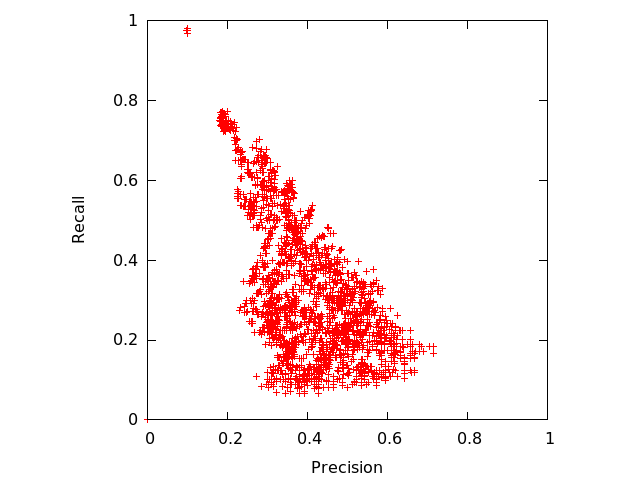}
\end{subfigure}~
\begin{subfigure}[t]{0.4\textwidth} 
\includegraphics[width=1\textwidth]{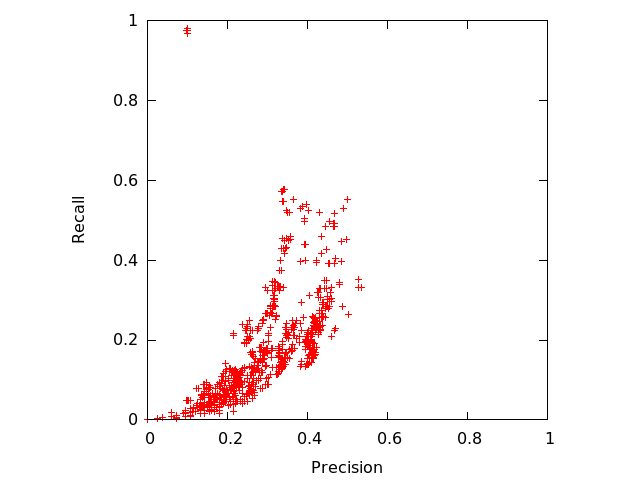}
\label{fig:abs_tuning}
\end{subfigure}~
\caption{Tuning absolute and relative thresholds $T$, text versus location weighting $r$ and population scaling exponent $\alpha$. Each point corresponds to the average precision and recall over 15 days for a different triple of $r,\alpha,T$. Left: Using relative floodiness. Right:  Using absolute floodiness.}
\label{fig:tuning}
\end{figure}

Fig \ref{fig:tuning} shows the average precision and recall for different parameter sets. Recall may be much more important than precision
i.e. it's better to get a false alarm than to have no warning before
a disaster. This is partly captured by the $F_{\beta}$ score, 
\begin{align}
F_{\beta}=\left(1+\beta^{2}\right)\frac{\text{precision}\times\text{recall}}{\beta^{2}\text{precision}+\text{recall}}
\end{align}
which is roughly a measure of accuracy given that recall is $\beta$
times as important as precision. The precision and recall depend more sensitively on $\alpha$ and $T$ than $r$. We only show the best parameter sets in Table \ref{tab:params}, however many other triples of ($r$, $\alpha$, $T$) are quite close or equivalent.

\begin{table}[!ht]
\centering
\caption{ 
{\bf Precision, recall and parameter set obtained by maximising $F_\beta$ scores using absolute and normalised floodiness.  }}
\begin{tabular}{ |c|c|c|c|c| }
\hline
$\beta$	& Max $F_\beta$ & Precision & Recall & $(r, \alpha, T)$ \\ \hline
\multicolumn{5}{|l|}{\bf Relative}\\ \hline
1 & 0.47 & 0.45 & 0.54 & $(\geq 2, 0.35, 0.25)$ \\ \hline
2 & 0.52 & 0.28 & 0.70 & $(\geq 1, 0.15, 0.1)$ \\ \hline 
\multicolumn{5}{|l|}{\bf Absolute}\\ \hline
1 & 0.53 & 0.50 & 0.55 & $(\geq 2, 0.4, 0.075)$ \\ \hline
2 & 0.54 & 0.50 & 0.55 & $(\geq 2, 0.4, 0.075)$ \\ \hline
\end{tabular}
\label{tab:params}
\end{table}

\begin{figure}[!ht]
\centering 

\begin{subfigure}[t]{0.4\textwidth} 
\includegraphics[width=\textwidth]{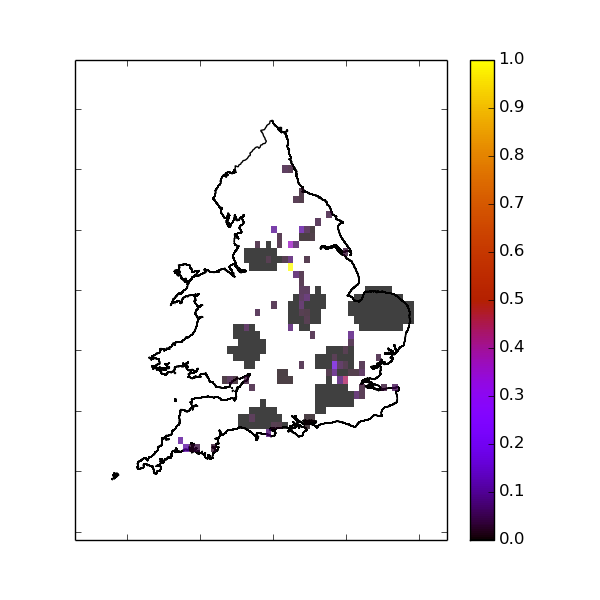}
\end{subfigure} ~
 \begin{subfigure}[t]{0.4\textwidth} 
 \includegraphics[width=\textwidth]{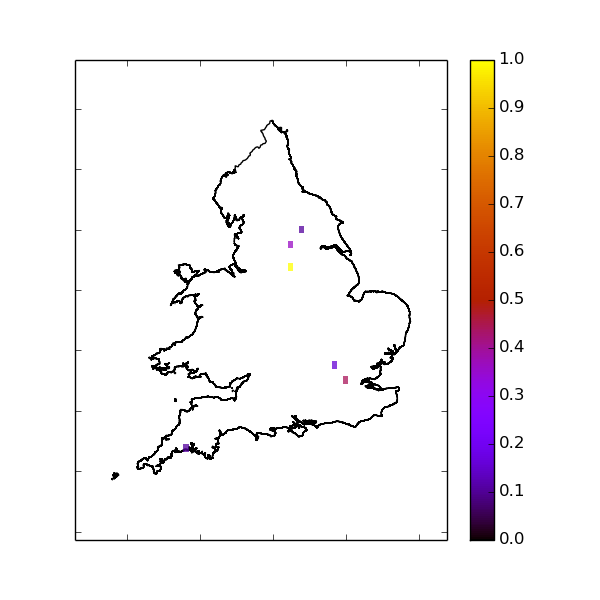}
\end{subfigure}

\begin{subfigure}[t]{0.4\textwidth}
\includegraphics[width=\textwidth]{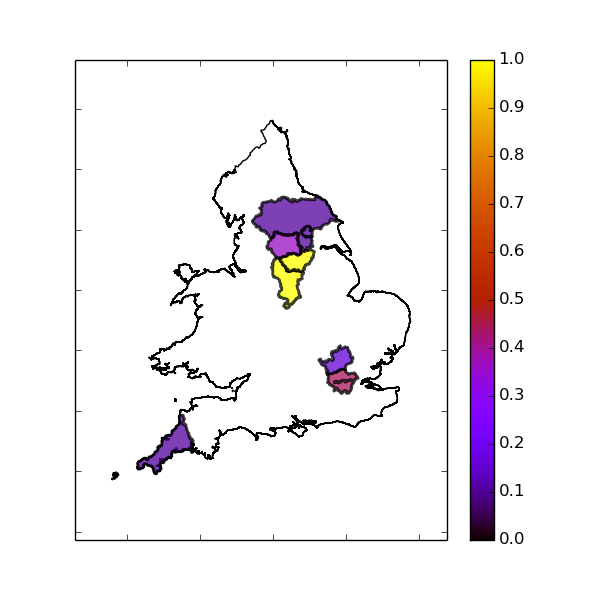}
\end{subfigure} ~
\begin{subfigure}[t]{0.4\textwidth}
\includegraphics[width=\textwidth]{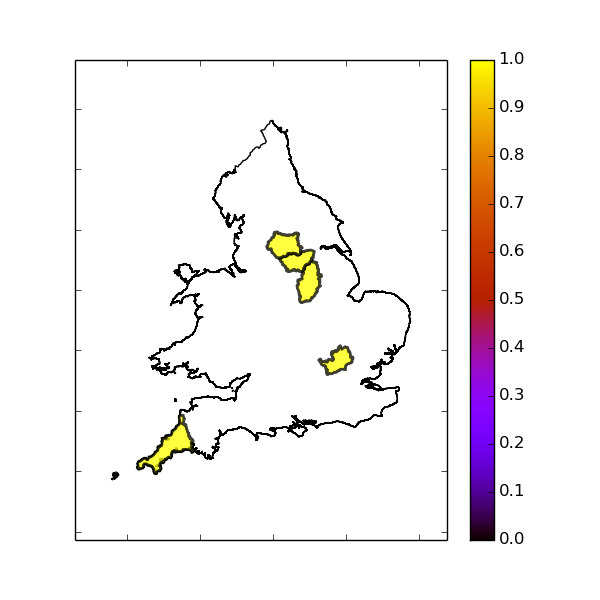}
\end{subfigure} 

\caption{Flood map generated by twitter converted into FFC format for validation. White indicated no tweets. Colour bar units are relative floodiness. Top Left: Floodiness grid ($64\times64$) over England and Wales on 28/10/2015
using $(r,\alpha)=(1.0, 0.15)$. Top Right: Showing only grid squares above threshold $0.1$. Bottom Left: Counties with floods on 28/10/2015 according to Twitter. Bottom Right: Counties with floods on 28/10/2015 according to the FFC, with $g_{h}$ set to 1 for flooded counties.}\label{fig:twitter_grid}

\end{figure}

Fig \ref{fig:twitter_grid} shows an example of the grid constructed
from one day, 28/10/2015, of flood tweets for one parameter set. We collected 48886 tweets in total for this day, of which only 302 passed all of the filters. Of these only 7 were geotagged while 222 had some form of location information in either the message text or the location field. 
There is reasonable agreement here at the level of counties, however Fig \ref{fig:twitter_grid} clearly shows that the twitter data is much higher resolution than county level. Counties have very variable sizes and populations, and the population within a county
is likely to be heavily concentrated in cities and towns - which are
the places where floods generate impacts that can be socially sensed most easily.
The higher resolution grid is more useful and more accurate and we
only transform it into a map of counties for verification purposes.

Though likely to be more accurate than Twitter data, the FFC data
we have is also not complete. Inspection suggests that many of our apparent false
positives, using the FFC data as truth, represent real flood events. e.g. 
\begin{verbatim}
`Tidal #Flooding @WarehamDorset this morning'
`M20 J2 London bound entry slip is closed due to flooding'
`@LeicsCountyHall, you need a highways team out urgently to Clarence St., 
Loughborough as gullies on both side are blocked and road is flooded'
\end{verbatim}
are all missing from the FFC data but seem likely to be reports of real floods. This suggests that our estimates of recall and especially precision may be too low and social sensing may be more accurate than suggested by comparison with FFC data. We suggest that the parameters $\alpha,T$ be changeable
at the user's discretion. The parameter $r$ is difficult to change
(requiring recomputing all the polygon overlaps) and but does not
affect precision and recall as much as $\alpha$ and $T$, therefore
we suggest fixing it at some $r \geq 1$. Lower $T$ improves recall but generates
more false alarms, higher $\alpha$ emphasises rural floods more than
cities. 

By showing that very high recall is possible we have demonstrated that we can reproduce the validation data using Twitter alone. The lower precision indicates some false positives but, by manual examination of the Tweets, shows us that we are detecting a lot of real floods not present in the validation dataset. Thus the socially sensed flood map provides a more complete record of historical floods than the FFC data. For verifying forecasts, showing every flood that occurred may be desirable. However for other uses - emergency response, detecting transport disruption, calculating insurance premiums - floods occurring in populous areas are of greatest interest and the threshold should be chosen differently. Thus we show the results for a range of $\alpha$, $T$ values and suggest that the end user find the best parameter set for their own purposes.

We also note that many false positives are references to historical
floods e.g. 
\begin{verbatim}
`Flood-damaged Leeds museum to re-open'
\end{verbatim}
which were not caught by our text filters. Tweets mentioning historical
floods may be of interest in analysing flood impact - more impactful
floods will have longer `tails' as people continue to talk about their
effects. This is the subject of ongoing work.

\section{Discussion}
\label{sec:discuss}

We have found that social sensing of floods is possible. The very
low volume of geo-tagged tweets makes location inference a necessity and
the accuracy of the location inference is crucial for the accuracy of the flood detection. We have outlined one location inference method here, which closely follows \cite{schulz:2013}, though we added some filtering steps to quickly localise tweets in the target area. We could imagine more elaborate systems e.g. where a user's friends contribute to their localisation \cite{rahimi:2015} or where users who consistently contribute accurate information are weighted more highly. We are actively pursuing these paths. Furthermore other major events will likely generate significant discussion on social media e.g. extreme wind, heat or cold; storms and hurricanes; earthquakes and very similar methods could be applied to all these cases. 

This approach does have biases, for example, we have more Tweets from areas with higher population density. We try to correct for this using the $\alpha$ parameter, but there is also likely to be
a demographic bias, since the Twitter user-base is not representative of the population 
in general. Other factors may also affect the volume of tweets collected e.g. the time of year (Christmas) or time of day (8am on a weekday) of the flood peak; how surprising the
the flood is or if the flood is covered by national news. These biases are an issue for
forecast verification and our relevance filter and population scaling attempt to correct for 
them. From the point of view of `live' flood tracking however, Twitter data provides low cost,
and highly relevant information - since people only tweet about things which are of interest!

We are currently working with the FFC to use social media data to improve their validation datasets and replace the current manual process of scouring local and national news for flood coverage. This should give flood forecasters richer and more reliable verification data to work with when improving their predictions. We are also working on a web based version of this code which is easier to interact with, with the aim of presenting it to emergency services, train and bus companies, councils or any other organization interested in assessing the historical impact of floods. A real time version, accessible on the web, would also be of interest e.g. for individuals worried about transport disruption. For both of these projects we must be careful to abide by Twitter's terms of service as well as privacy laws \cite{GDPR}.

Social media fills a gap by providing real time data about extreme events which is difficult to obtain otherwise. It provides excellent validation data for forecast models as well as live, on-the-ground updates during extreme events. As it stands the method presented in the paper is a proof of concept - social sensing using Twitter can provide accurate and useful information about flooding - and the method presented should already be useful for groups interested in verifying flood forecasts, managing flood responses in real time or tracking the social impact of floods.

\nolinenumbers

\section*{Acknowledgements}

The authors would like to thank the Met Office and UK Environment Agency Flood Forecasting Centre for providing historical data on floods in the UK.

\end{document}